\documentclass{vgtc}                          % final (conference style)
\graphicspath{{figures/}{pictures/}{images/}{./}} % where to search for the images

\usepackage{times}                     % we use Times as the main font
\usepackage{graphicx}
\usepackage{wrapfig}
\usepackage{amsmath}
\usepackage{amssymb}

\usepackage{multirow}

\usepackage[most]{tcolorbox}
\usepackage{lipsum}

\usepackage{tabu}                      % only used for the table example
\usepackage{booktabs}                  % only used for the table example
\usepackage{lipsum}                    % used to generate placeholder text
\usepackage{mwe}                       % used to generate placeholder figures

\usepackage{mathptmx}                  % use matching math font

\newcommand{\ssection}[1]{\noindent {\bf #1} }

\onlineid{1073}

\vgtccategory{Research}

\vgtcinsertpkg

\title{Crafting Dynamic Virtual Activities with Advanced Multimodal Models}

\author{Changyang Li\thanks{e-mail: changyang.li@goertekusa.com}\\ %
        \scriptsize Goertek Alpha Labs %
\and Qingan Yan\thanks{e-mail: qingan.yan@goertekusa.com}\\ %
     \scriptsize Goertek Alpha Labs %
\and Minyoung Kim\thanks{e-mail: mkim229@gmu.com}\\ %
     \scriptsize George Mason University %
\and Zhan Li\thanks{e-mail: zhan.li@goertekusa.com}\\ %
     \scriptsize Goertek Alpha Labs %
\and \begin{tabular}[t]{c@{\hspace{2em}}c}
    Yi Xu\thanks{e-mail: yi.xu@goertekusa.com} & Lap-Fai Yu\thanks{e-mail: craigyu@gmu.com} \\
    \scriptsize Goertek Alpha Labs & \scriptsize George Mason University \\
    \end{tabular}%
}

\teaser{
  \centering
  \includegraphics[width=\linewidth]{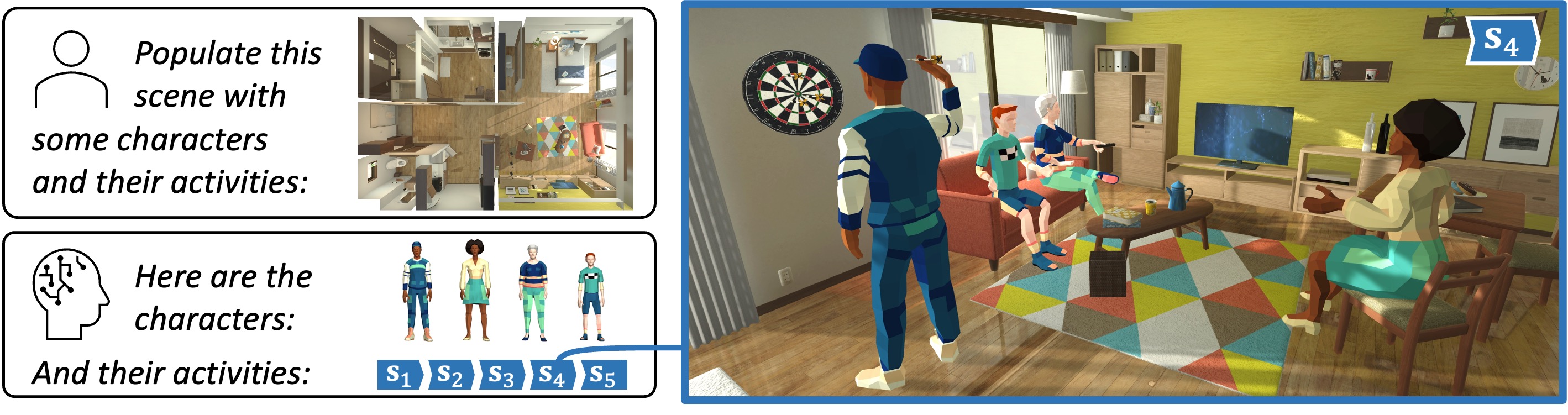}
  \caption{Given a virtual environment as the input, our method leverages multimodal large language models (MLLMs) to automatically generate contextually realistic character interactions. $\mathbf{s}_t$ denotes scene snapshots at discrete keyframe $t$. $\mathbf{s}_4$ here features two characters engaging in a conversation, an elderly female watching TV, and another character participating in a game of darts.}
  \label{fig:teaser}
}

\abstract{
    In this paper, we investigate the use of multimodal large language models (MLLMs) for generating virtual activities, leveraging the integration of vision-language modalities to enable the interpretation of virtual environments. Our approach recognizes and abstracts key scene elements including scene layouts, semantic contexts, and object identities with MLLMs’ multimodal reasoning capabilities. By correlating these abstractions with massive knowledge about human activities, MLLMs are capable of generating adaptive and contextually relevant virtual activities. We propose a structured framework to articulate abstract activity descriptions, emphasizing detailed multi-character interactions within virtual spaces. Utilizing the derived high-level contexts, our approach accurately positions virtual characters and ensures that their interactions and behaviors are realistically and contextually appropriate through strategic optimization. Experiment results demonstrate the effectiveness of our approach, providing a novel direction for enhancing the realism and context-awareness in simulated virtual environments.
} % end of abstract


\begin{document}


\firstsection{Introduction}
\label{sec:intro}

\maketitle

The generation of virtual environments and the simulation of human activities within these spaces are increasingly important in computer graphics, computer vision, and robotics. Recent advances, including the availability of large-scale datasets and sophisticated deep generative models, have significantly improved the capability to produce lifelike human-scene interactions.

However, existing approaches face several challenges. Many current methods primarily address the placement of virtual characters~\cite{zhang2020generating,zhang2020place,hassan2021populating,zhao2022compositional} based on low-level geometric and physical constraints, without adequately incorporating high-level contextual information such as the nature of the activities performed. Additionally, although task- or instruction-specific motion synthesis methods~\cite{cao2020long,starke2019neural,wang2021synthesizing,wang2021scene,hassan2021stochastic,xu2023interdiff,hassan2023synthesizing,huang2023diffusion,wang2022towards,zhao2023synthesizing} have emerged, they often overlook the dynamic interplay among multiple characters and their interactions with surrounding objects, limiting realism and scenario complexity. Addressing these limitations demands methods capable of interpreting and leveraging the semantic context and spatial layout within virtual scenes, and correlating such clues with human activities and behaviors.

Recent advances in language models, particularly those employing deep learning and transformer architectures~\cite{vaswani2017attention} and supporting multimodality~\cite{radford2021learning,ramesh2022hierarchical,alayrac2022flamingo,li2022blip}, have revolutionized the field of natural language processing and computer vision. These models demonstrate remarkable capabilities in understanding and generating complex language and vision contexts. The integration of linguistic strengths into multimodal large language models (MLLMs) presents new opportunities for enhancing virtual activity generation: Leveraging advanced linguistic reasoning and visual understanding, MLLMs can interpret environments, understand character dynamics, and generate contextually relevant interactions.

Motivated by these developments, our research aims to utilize the potential of MLLMs in generating virtual activities, especially for multi-character human-scene interactions involving multiple characters. By integrating vision-language modalities through a Layout Chain-of-Thought (L-CoT) prompting strategy, we enable MLLMs to interpret virtual environments visually, understand semantic contexts, infer spatial relationships, and perform logical reasoning when generating human activities. We propose a structured framework that explicitly defines virtual activities in terms of character poses, positional references, and detailed interactions on discrete keyframes. Furthermore, our method employs an efficient optimization with a Markov chain Monte Carlo (MCMC) sampling process to precisely position virtual characters, ensuring natural, contextually consistent interactions aligned with their generated high-level activity descriptions.

Generating contextually meaningful virtual activities holds great potential in numerous virtual and augmented reality (VR/AR) application scenarios, including training virtual agents for social scenarios, creating enhanced immersive VR and AR experiences, facilitating advanced human-robot interactions, and supporting virtual storytelling and content creation.

\begin{figure*}[t]
     \centering
     \includegraphics[width=0.98\textwidth]{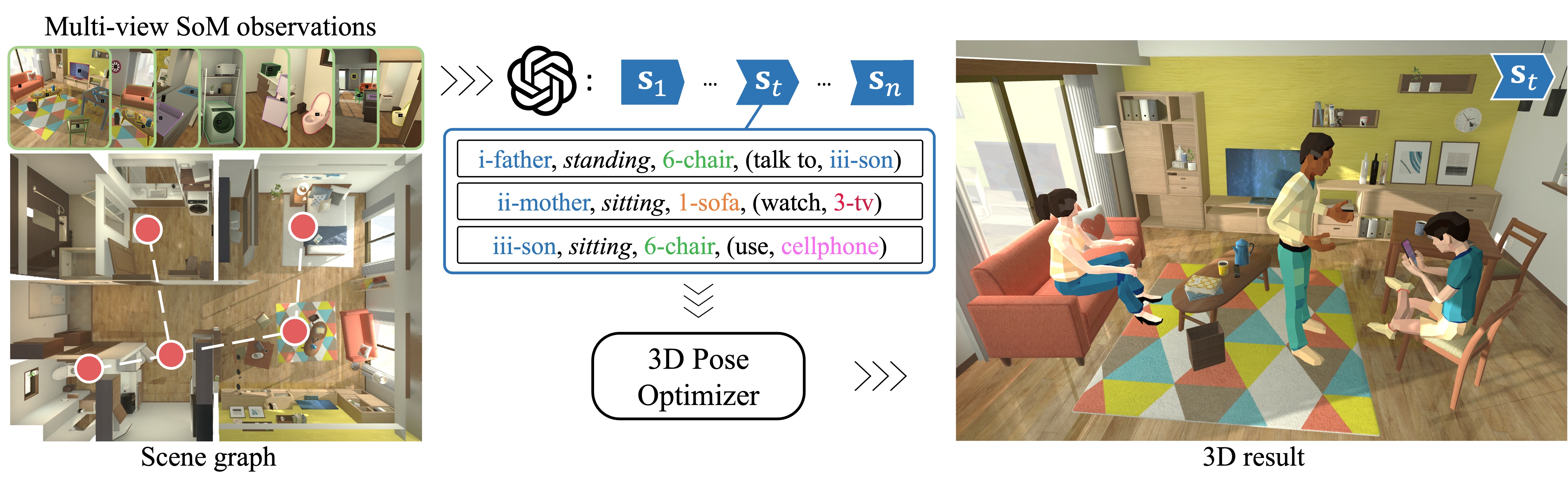}
     \vspace{-5mm}
     \caption{An overview of our approach. We first capture multi-view Set-of-Mark (SoM)~\cite{yang2023set} observations of the scene, enabling the MLLM to construct an area-based scene graph and interpret spatial layouts. Next, the MLLM generates detailed descriptions of virtual activities at discrete keyframes, specifying characters’ poses, positional references, and interactions. Finally, a 3D pose optimizer positions the virtual characters in the scene. The example shows generated activity descriptions for a selected keyframe ($\mathbf{s}_t$) and their corresponding 3D realization.} 
     \vspace{-4mm}
     \label{fig:overview}
\end{figure*}

The major contributions of our work include the following: 
\begin{itemize}
    \item Proposing a multimodal reasoning approach using MLLMs for detailed interpretations of virtual environments;
    \item Formulating a structured representation of virtual activities to effectively prompt MLLMs for context-aware, adaptive, and realistic generation of multi-character interactions; 
    \item Devising an efficient optimization strategy for realistic and scalable character placement in virtual scenes.
\end{itemize} 

\section{Related Work}
\label{sec:related}

\subsection{Scene-Adaptive Activity Generation}

The generation of scene-aware virtual human behaviors is pivotal for creating realistic virtual environments. A key challenge in this domain is the adaptive generation of virtual human poses within input scenes. Savva et al. explored the relationship between the geometries and functionalities of 3D environments and the corresponding human poses and object arrangements, leveraging observations from human-scene interactions~\cite{savva2014scenegrok,savva2016pigraphs}. Some recent methods utilizing deep generative models have focused on posing virtual humans by considering factors such as geometry, semantics, affordances, and contact points~\cite{zhang2020generating,zhang2020place,hassan2021populating,zhao2022compositional}.

Beyond static snapshots of human poses, the generation of scene-aware human motion considers temporal dynamics and the interactions between humans and their environments, guided by specific tasks~\cite{bai2012synthesis,agrawal2016task,pirk2017understanding,lee22control}, action labels~\cite{guo2020action2motion,petrovich2021action} or the surrounding environments~\cite{cao2020long,starke2019neural,wang2021synthesizing,wang2021scene,hassan2021stochastic,xu2023interdiff,hassan2023synthesizing,huang2023diffusion,wang2022towards,zhao2023synthesizing}. There is also a growing interest in integrating virtual humans into real-world environments for augmented reality applications, highlighting the need for adaptive and interactive behaviors~\cite{tahara2020retargetable,li2022interactive,kari2023scene}.

While those methods mainly concentrate on the motion and behaviors of individual characters, our research shifts focus towards scenarios involving multi-character interactions within scenes. Some related researches that study detailed multi-character interactions~\cite{kim2021interactive,zhang2023simulation,li2024two,wang2024intercontrol}, but overlook the correlations with the surroundings. Crowd simulations provide strategies for driving large groups of characters in environmental contexts; however, these approaches typically do not account for intricate interactions among characters or the higher-level interpretation of the activities being performed~\cite{paris2007pedestrian,lee2018crowd,zhao2018clust,charalambous2023greil}. Li et al.~\cite{li2023generating} addressed a closely related problem by proposing the synthesis of activity snippets that depict sequenced multi-character and multi-object interaction scenarios. In comparison, our work focuses on adapting virtual activities to given environments, rather than jointly creating virtual environments. Whereas activity snippets~\cite{li2023generating} may face scalability issues in larger group settings without extensive training data, our approach leverages the knowledge embedded in MLLMs to facilitate large-scale, multi-character interactions within extensive environments, demonstrating a novel adaptation of these models in generating contextually rich and dynamic virtual scenes.

\subsection{Language Models for Activity Generation}

Advancements in language models have significantly influenced research directions such as activity generation and behavior modeling. For example, LLMs can simulate human behaviors within interactive virtual sandbox environments\cite{park2023generative,li2024evolving}. More recent research has investigated employing language models to generate detailed human motion directly from textual inputs\cite{ghosh2021synthesis,athanasiou2022teach,zhong2024smoodi,li2024unimotion}, or to produce motion sequences adapted to specific scenes and contexts~\cite{xiao2023unified,wang2022humanise,wang2024move,yi2024generating,jiang2024autonomous}. Existing approaches also use scene graphs that capture rich scene context to guide LLMs through high-level activity or task planning~\cite{gorlo2024long,li2025x}. 

Recent models with vision-language modalities demonstrate great power in various directions, like 3D scene understanding~\cite{huang2023chat,huang2023embodied,hong20243d,chen2024ll3da,chen2024grounded,gu2024conceptgraphs} and VR/AR world enhancement~\cite{chen2025llmer,lee2025imaginatear}. To realize scene-aware activity generation, adopting vision-language models is critical to understanding scene layouts and semantics. These works provide insights into using MLLMs for an enriched understanding and generation of activities within 3D environments. Our approach similarly employs MLLMs to grasp the environmental cues of the target scenes, facilitating the creation of virtual activities that are both contextually aware and semantically compatible.

\section{Overview}

\autoref{fig:overview} shows an overview of our approach. It begins with enabling multimodal large language models (MLLMs) to interpret virtual 3D environments through two types of scene clues: area descriptions and multi-view image observations (Section~\ref{sec:activity}). To facilitate precise visual grounding, we adopt the Set-of-Mark (SoM)~\cite{yang2023set}, which explicitly labels object identities. We then employ a Layout Chain-of-Thought (L-CoT) prompting strategy, guiding MLLMs through explicit reasoning steps to infer spatial adjacencies between areas and construct area-based scene graphs.

Using the interpreted environmental context, we then prompt MLLMs to generate dynamic, context-aware virtual activities involving multiple characters (Section~\ref{sec:activity_descriptions}). Each virtual activity is structured as a sequence of discrete keyframes, where each keyframe describes character poses, positional references, and detailed interactions with objects or other characters.

Finally, to realize these high-level symbolic descriptions within the 3D virtual environments, we apply an efficient optimization strategy (Section~\ref{sec:optimization}). This procedure leverages Markov chain Monte Carlo (MCMC) sampling to precisely position characters. Animation clips corresponding to the generated descriptions are applied to produce realistic animations of the virtual activities.

\section{Understanding 3D Environments Using Vision-Language Models}

In the initial phase, we focus on enabling vision-language MLLMs to understand virtual 3D scenes using combined visual and linguistic information. We specifically use GPT-4o for all results. 

\subsection{Scene Clues Preparation}

We provide MLLMs two types of scene clues: area descriptions and multi-view image observations. These inputs help MLLMs interpret scene layouts, spatial relationships, object shapes, functionalities, and affordances. Area descriptions are later converted to area-based scene graphs by the MLLMs.

\ssection{Area-based Scene Graph.} Scene graphs provide abstracts of spatial layouts to MLLMs. In this work, we define simplified scene graphs conveying only basic structural information, omitting specific numeric or specialized attributes: Instead of using a highly structured scene graph representation with comprehensive details encoded, which demands careful, object-specific design, we leave such details to be understood using vision-language models without requiring explicit, domain-specific attribute definitions. 

Specifically, we first group scene objects into distinct clusters called \textbf{scene areas}, considering distances and visibilities. Pairwise visibility is determined if the two objects are visible to each other (e.g., not separated by a wall). An area is modeled as a single node in the scene graph. Scene area descriptions are composed into json format as inputs for MLLMs.

To connect these area nodes to build a scene graph, one could
apply standard minimum spanning tree (MST) algorithms such as Prim's~\cite{prim1957shortest} or Kruskal's~\cite{kruskal1956shortest}. However, we found that allowing MLLMs to perform this task, guided by visual clues from multi-view images, produces more accurate and spatially coherent results. Given a fixed scene graph with abstract scene information, MLLMs might overlook details that need to be inferred through visual clues, and misinterpret spatial layouts. For example, identical graph distances could correspond to different actual spatial distances. Our ablation tests in Section~\ref{sec:ablation} further support this observation. Thus, we defer this step to be finished by MLLMs as an explicit reasoning step after interpreting the visual clues from multi-view images. An example scene graph is visualized in~\autoref{fig:overview}.

\ssection{Multi-view Image Observations.} We feed multi-view scene images to MLLMs, which help them recognize scene objects, their functionalities, and affordances. All images are captured by a virtual camera that navigates randomly to ensure comprehensive coverage of the scene. This ensures diverse view angles and improves the MLLMs’ spatial reasoning and contextual understanding.

However, aligning recognized objects precisely with their scene identities remains challenging for current MLLMs. To address this, we employ the Set-of-Mark (SoM)~\cite{yang2023set} representations, where each object is clearly marked in images with a unique ID label, facilitating consistent identification across different views. The marks are based on instance segmentations sourced from the virtual 3D scenes. Each object within the scene is assigned a unique ID for subsequent reference when the view changes. In practice, we filter out marks based on size, distance, visibility, and overlap criteria, removing those that are too small, overly distant, significantly outside the field of view, or heavily occluded. By default, we remove marks smaller than $4\%$ of the image area, further than $10m$ from the camera, partially out of view by more than $20\%$, or substantially overlapped with other marks. In real-world applications, these segmentations for markings could be obtained using state-of-the-art segmentation methods like the Segment Anything (SAM) family~\cite{kirillov2023segment,ravi2024sam}. To maintain consistent object identities across different views, 2D segmentation results can be projected onto the 3D digital twin. Alternatively, 3D segmentations can be directly applied, with corresponding 2D segmentation masks rendered for each view.

\begin{figure}[t]
     \centering
     \includegraphics[width=0.42\textwidth]{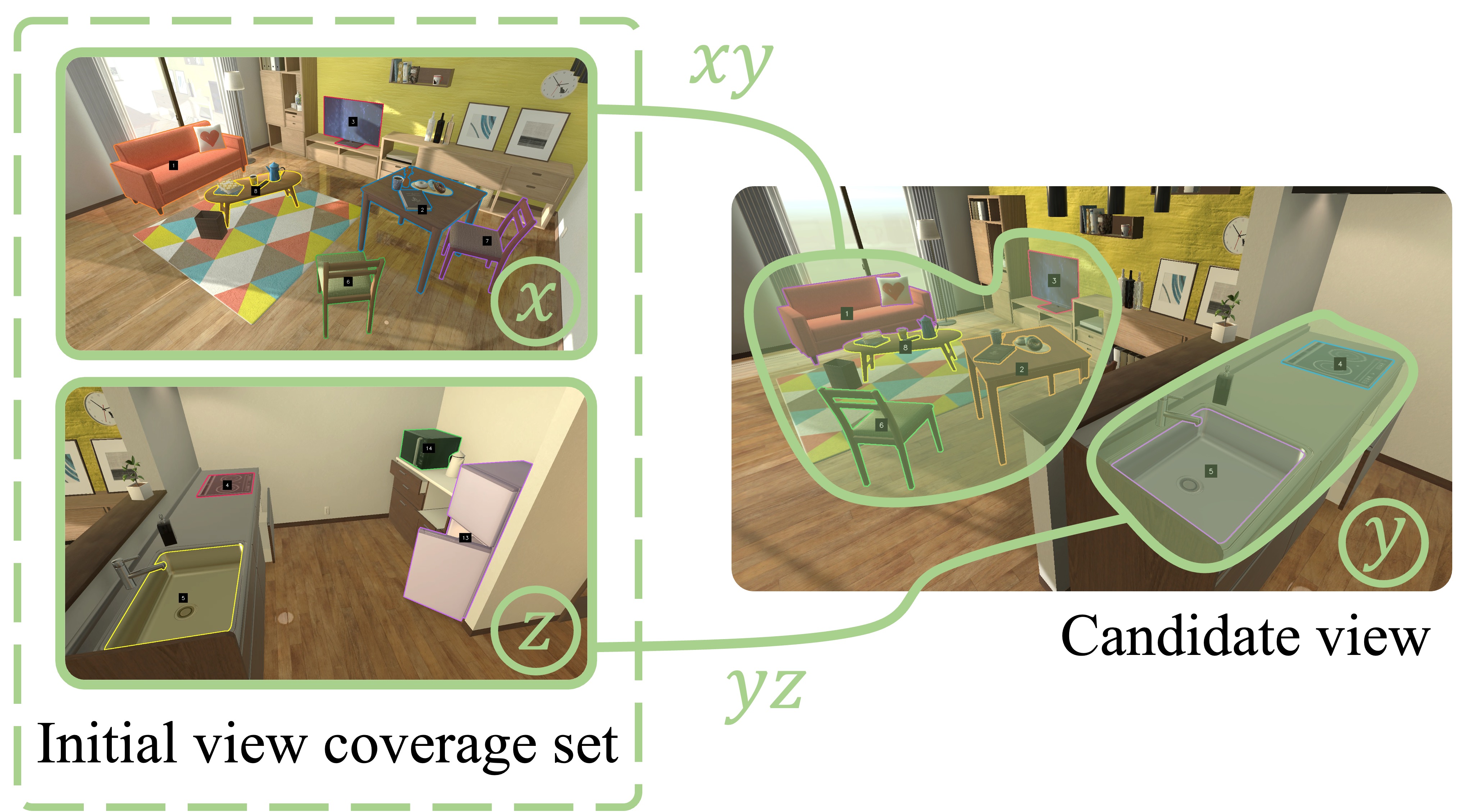}
     \vspace{-3mm}
     \caption{An illustration of maintaining connectivity during the selection of multi-view observations. The initial coverage set includes views $x$ and $z$ covering distinct scene objects. A candidate view $y$ is selected to connect $x$ and $z$ by capturing shared objects.}
     \vspace{-4mm}
     \label{fig:connectivity}
\end{figure}

To reduce information redundancy, we keep a minimum subset of views that still cover all needed scene objects and preserve necessary scene clues. Starting from an empty set, we iteratively select images that cover the largest number of not-yet-covered objects until all objects are covered, and get an initial view coverage set.

Another important consideration is to ensure view traversability, that the views form a connected graph under the rule ``two views are adjacent if they share at least one object''. This is an essential consideration for the Layout Chain-of-Thought (L-CoT) we use to guide MLLMs in understanding the environments, of which details will be discussed later. Continuing from the initial coverage set, we keep selecting additional views, prioritizing views that connect two or more previously disconnected components, meanwhile maximizing the number of shared objects with the images in those components, as depicted in ~\autoref{fig:connectivity}. Note that traversability cannot be guaranteed (e.g. when an area is in a separate room), and in such cases we process the remaining unconnected areas with MST.

\subsection{Layout Chain-of-Thought Prompting} 

\begin{figure}[t]
     \centering    \includegraphics[width=0.48\textwidth]{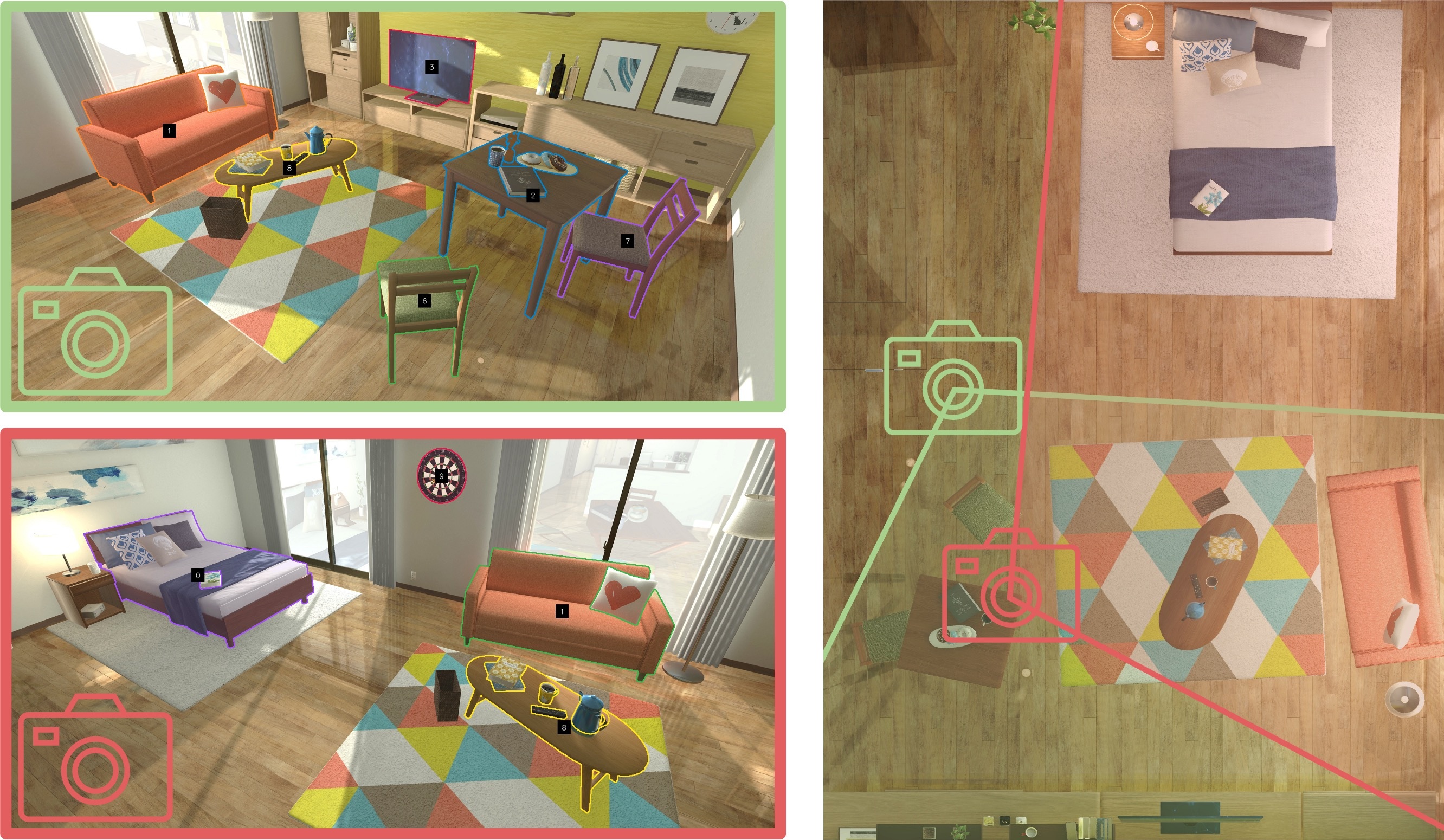}
     \caption{The proximity between areas captured by two camera views is determined following the Layout Chain-of-Thought (L-CoT) prompting, given their shared common objects.}
     \label{fig:cot}
\end{figure}

Aligning with the principles behind Chain-of-Thought~\cite{wei2022chain}, which simplifies complex queries into manageable steps, we propose Layout Chain-of-Thought (L-CoT) for layout understanding. This involves guiding MLLMs through a sequence of reasoning steps: We first require MLLMs to explicitly identify objects associated with their areas according to the area descriptions. Next, MLLMs should list shared objects appearing in different views as an intermediate reasoning step, and further determine the views' spatial adjacencies. Areas' adjacencies could be reasoned if one view contains objects from multiple areas, or reasoned from multiple views: For instance, while each view shown in~\autoref{fig:cot} focuses on a distinct small area of the scene, the shared objects (sofa and tea table) suggest the proximity between these areas. Finally, MLLMs should use the reasoned adjacencies as edges to complete the area-based scene graph, given areas defined before as nodes.

L-CoT explicitly guides MLLMs through intermediate reasoning steps to interpret spatial layouts. This reasoning approach improves the model’s ability to accurately infer spatial relationships, thereby enhancing the realism of subsequent virtual activity generation, which also uses L-CoT and is discussed in Section~\ref{sec:activity}.

\section{Dynamic Virtual Activities}
\label{sec:activity}

In our work, a virtual activity is formally defined as a tuple $(\mathbf{C}, \tau)$, summarizing various components essential for activity representation in virtual environments. The set $\mathbf{C} = \{c\}$ represents the characters involved in the activity, with each character $c$ is identified by a unique ID and associated with a specific role, which could range from social roles such as a parent to professional roles like a chef. The snapshots $\tau = \{\mathbf{s}_t\}_{t=1,2,\dots,T}$ depicts a sequence of scene states across keyframes from $1$ to $T$, thereby discretizing the activity into manageable segments.

\subsection{Generating Activity Descriptions with MLLMs}
\label{sec:activity_descriptions}

A core aspect of our approach is the generation of snapshots $\tau$ of the activity using MLLMs. Each snapshot $\mathbf{s}_t$ provides all characters' states at a specific keyframe $t$, expanded into a series of detailed descriptions such that $\mathbf{s} = \{d\}$. A single description $d$ details an individual character's pose and ongoing activity at the given keyframe. We define the following attributes for structuring each description $d = (\alpha, p, r, (v, \beta))$:

\begin{enumerate}
    \item \textbf{Subject} $\alpha$, which is the focal character of the description.
    \item \textbf{Pose} $p$, which distinguishes among three fundamental poses: \textit{standing}, \textit{sitting}, and \textit{lying}. These poses cover a majority of life scenarios as the base, and actual poses can be further detailed with exact body parameters defined in Section~\ref{sec:optimization} and specific animation clips (e.g. \textit{sitting + talking}).
    \item \textbf{Positional Reference} $r$, which enhances layout comprehension and activity planning through MLLMs by anchoring characters to specific furniture or appliances corresponding to their poses (i.e. \textit{standing near}, \textit{sitting on}, and \textit{lying on}).
    \item \textbf{Interaction} $(v, \beta)$, which enumerates the interactions involving the described character. Each interaction is characterized by a verb $v$ (e.g., ``read'', ``talk to'') and a grammatical object $\beta$ (e.g., object\_10-book, character\_2) that $\alpha$ interacts with.
\end{enumerate}

This structured format leverages MLLMs' capabilities in interpreting and generating complex scenarios, thereby facilitating detailed and dynamic representations of virtual interactions.

\begin{figure*}[htbp]
    \centering
    \begin{tcolorbox}[colback=gray!15, colframe=gray!70, width=\textwidth, boxrule=0.5mm, arc=4mm, auto outer arc, boxsep=5pt]
        \textbf{System:} You are a visual assistant interpreting 3D scenes containing multiple objects. Inputs provided to you: 1. Scene areas (JSON), each including an ``objects'' list; 2. Multi-view scene images with unique object IDs labeled at object centers. 

You must explicitly perform two tasks below, following the exact steps:

Task I. Build a Scene Graph: 

Step 1. List objects grouped by their areas clearly from all provided images.

Step 2. Identify visual adjacencies between scene views based on objects appearing in multiple images from Step 1.

Step 3. Construct a Scene Graph (JSON), using areas as nodes, with an additional ``adjacent'' attribute, and connect nodes (areas) explicitly if their adjacencies are confirmed by shared objects. Clearly state reasons for each connection using evidence from previous steps.

Task II. Generate a Virtual Activity: 

Step 1. Decide the number of characters appropriate for the scene scale, and assign unique IDs and relevant roles (family/social/professional) based on scene context.

Step 2. Activity creation (sequence of keyframes): Each keyframe is a list of all characters' states at that time. A character's state is defined as: (ID, pose, reference, interaction), with ``ID'' being the assigned character ID, ``pose'' being ``standing'', ``sitting'', or ``lying'', ``reference'' being the object for ``standing nearby'' or ``sitting on'' or ``lying on'', ``interaction'' being a tuple (type, interactee) where ``type'' is the action performed (e.g., ``talk to'', ``use'') and ``interactee'' is the character/object involved in interaction.
Ensure continuous transitions between keyframe: Every change in a character's state triggers a new keyframe. When changing states, please clearly justify your reasoning. For movements to a new ``reference'' object, explicitly describe intermediate steps based on the scene graph and shared object adjacencies observed in the images.

Provide explicit, structured answers following these guidelines.

        \vspace{0.3cm}
        \textcolor{gray}{for \textcolor{blue}{sample} in \textcolor{blue}{fewshot\_samples:}}

        \vspace{0.2cm}
        \begin{tcolorbox}[colback=white, colframe=blue, width=\textwidth, boxrule=0.2mm, arc=2mm]
            \textbf{User:} \{``area\_0'' : \{``objects'': [``object\_0-sink'', ``object\_1-stove'', ``object\_2-cooktop'']\}, ``area\_1'' : \{``objects'': [``object\_3-bookshelf'', ``object\_5-bed'']\}, ``area\_2'' : \{``objects'': [``object\_4-tea\_table'', ``object\_6-chair'', ``object\_7-chair'', ``object\_8-table'', ``object\_9-standing\_lamp'', ``object\_10-tv'', ``object\_11-sofa'']\}\}

            \vspace{0.2cm}
            \noindent
            \includegraphics[width=0.24\textwidth]{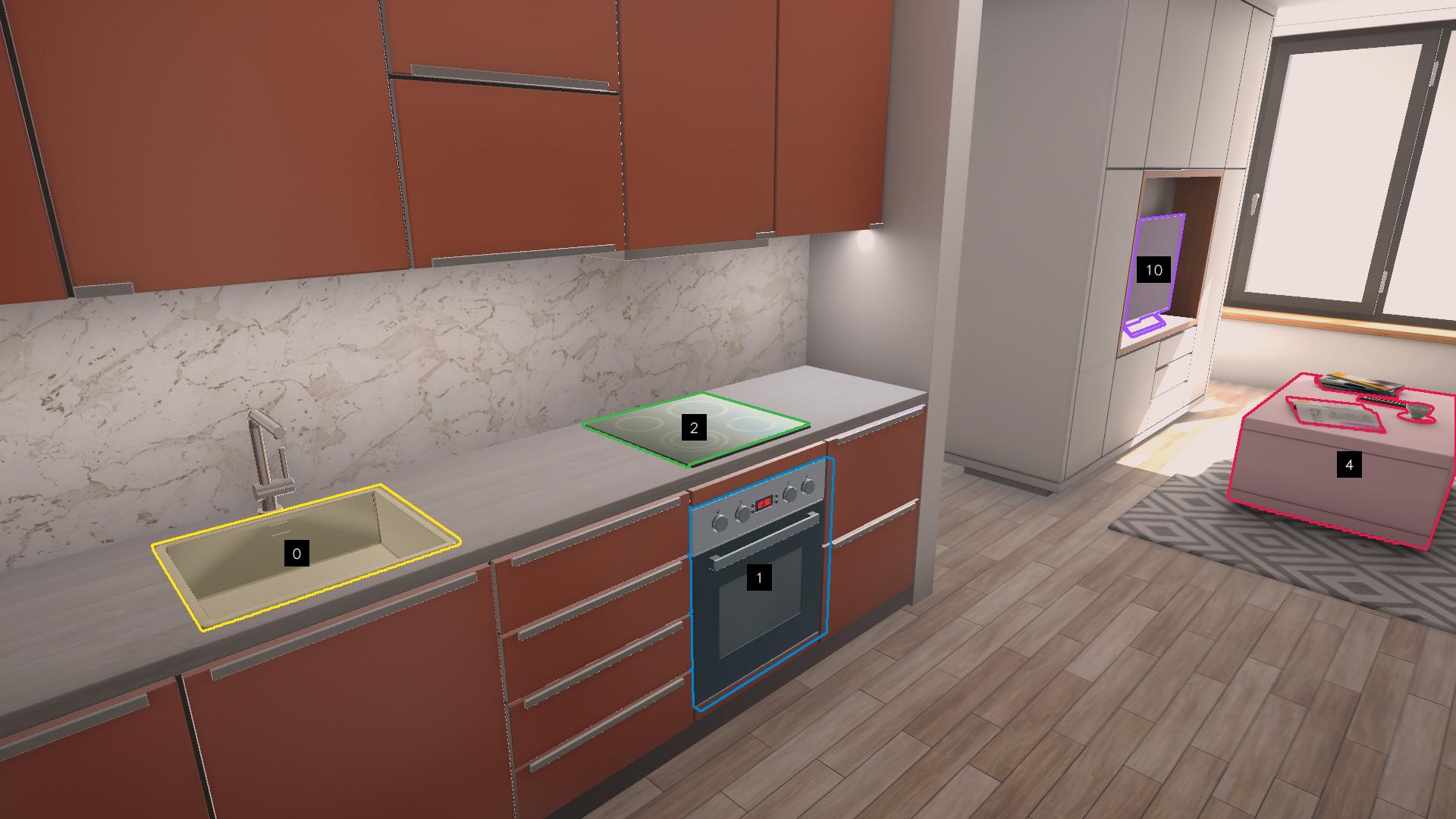}\hfill
            \includegraphics[width=0.24\textwidth]{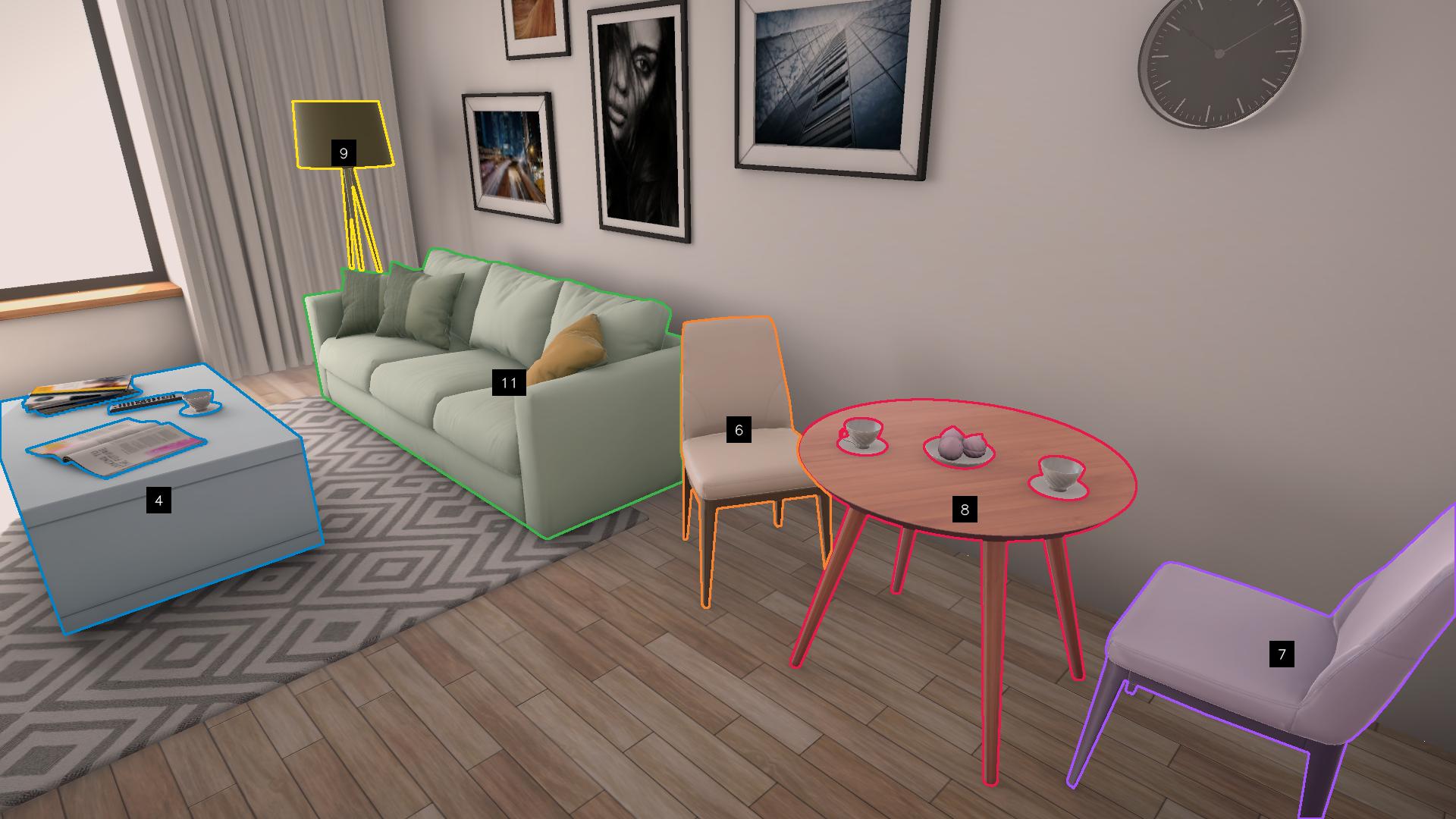}\hfill
            \includegraphics[width=0.24\textwidth]{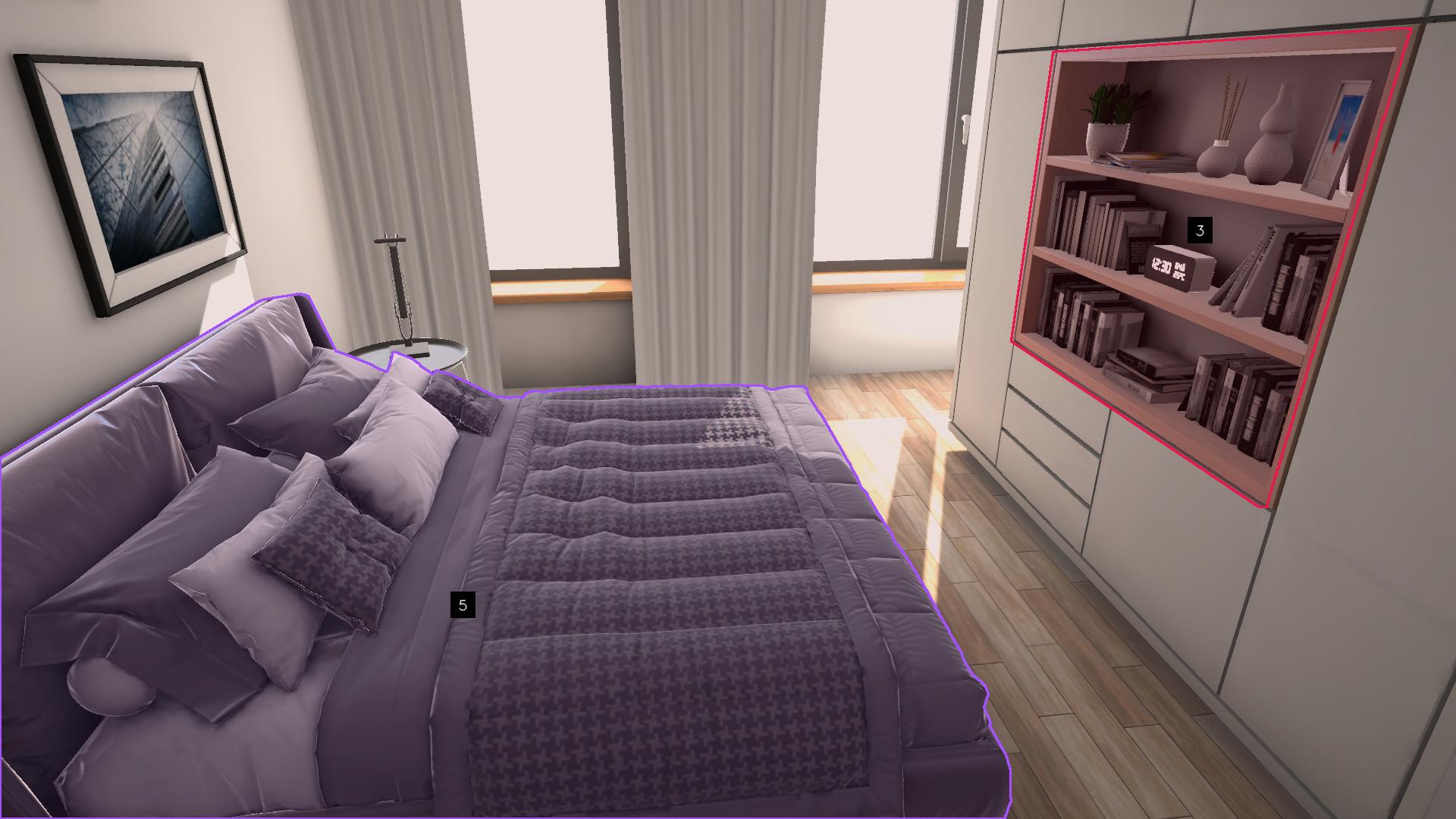}\hfill
            \includegraphics[width=0.24\textwidth]{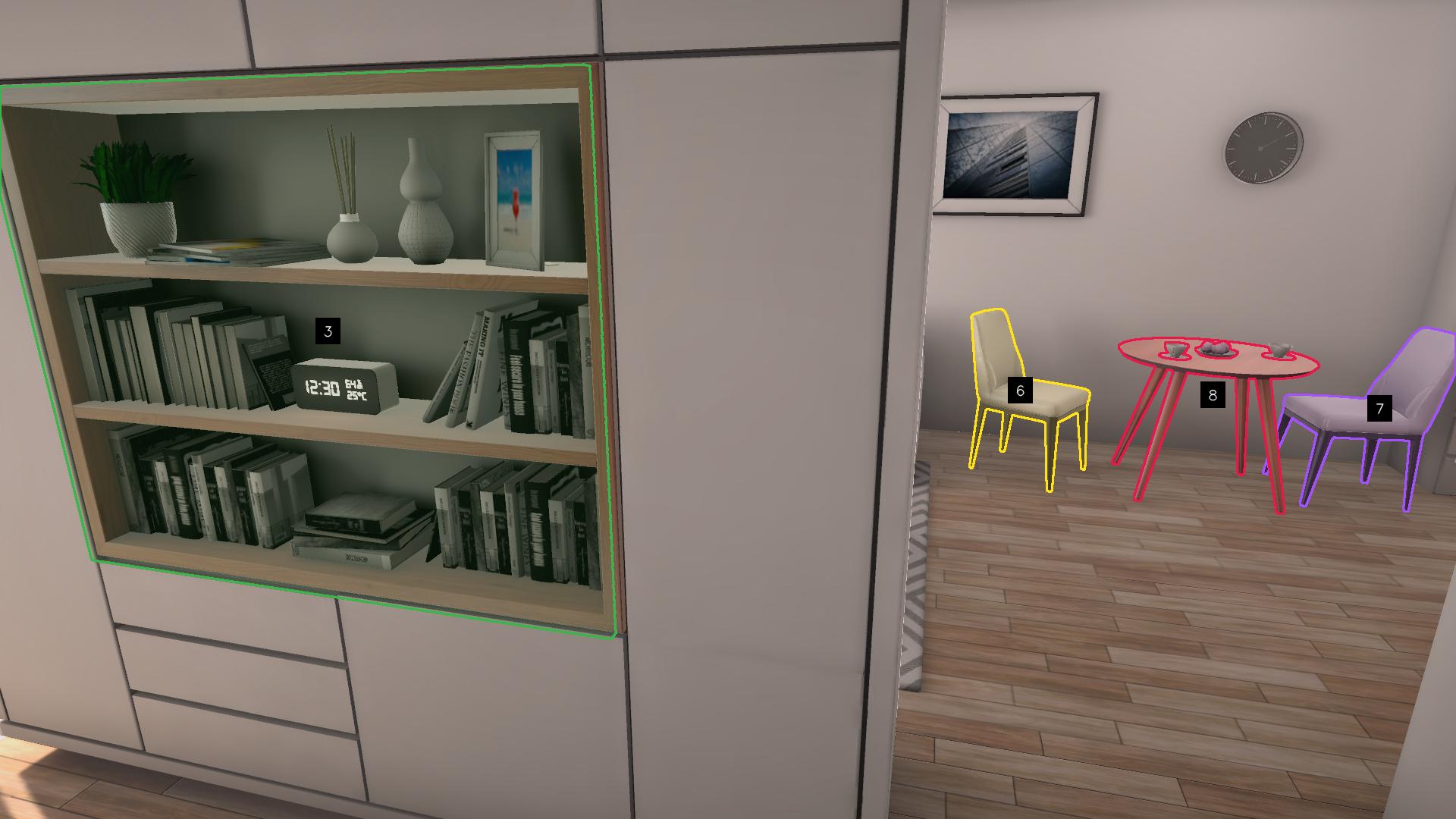}

            \vspace{0.2cm}
            \textbf{Assistant:} Task I.1: View 0: area\_0: [object\_0-sink, object\_1-stove, object\_2-cooktop], area\_2: [object\_4-tea\_table, object\_10-tv]; View 1: area\_2: [object\_4-tea\_table, object\_6-chair, object\_7-chair, object\_8-table, object\_9-standing\_lamp, object\_11-sofa]; View 2: area\_1: [object\_3-bookshelf, object\_5-bed]; View 3: area\_1: [object\_3-bookshelf], area\_2: [object\_6-chair, object\_7-chair, object\_8-table].

I.2: View 0 and View 1 are visually adjacent because they both include object\_4; View 2 and View 3 are visually adjacent because they both include object\_3; View 1 and View 3 are visually adjacent because they both include object\_6, object\_7 and object\_8.

I.3: View 1 contains area\_0: [object\_0, object\_1, object\_2], and area\_2: [object\_4, object\_10], thus area\_0 and area\_2 are adjacent; View 4 contains area\_1: [object\_3, object\_5] and area\_2: [object\_6, object\_7, object\_8], thus area\_1 and area\_2 are adjacent. Overall, the scene graph can be represented as:

\{``area\_0'' : \{``objects'': [``object\_0-sink'', ``object\_1-stove'', ``object\_2-cooktop''], ``adjacent'': [``area\_2'']\}, ``area\_1'' : \{``objects'': [``object\_3-bookshelf'', ``object\_5-bed''], ``adjacent'': [``area\_2'']\}, ``area\_2'' : \{``objects'': [``object\_4-tea\_table'', ``object\_6-chair'', ``object\_7-chair'', ``object\_8-table'', ``object\_9-standing\_lamp'', ``object\_10-tv'', ``object\_11-sofa''], ``adjacent'': [``area\_0'', ``area\_1'']\}\}

Task II.1: Based on the scale of the scene that looks like a studio apartment for one to two people to live in, I'll create two virtual characters: character\_0: Husband; character\_1: Wife.  

II.2: keyframe 0:  
(character\_0, sitting, object\_11-sofa, (read, newspaper)). \textcolor{gray}{Thoughts: character\_0 starts with reading newspaper.} (character\_1, lying, object\_5-bed, (sleep, none)). \textcolor{gray}{Thoughts: character\_1 starts with sleeping on the bed.}

keyframe 1:  
(character\_0's state does not change)  
(character\_1, standing, object\_0-sink, (get water, object\_0-sink)). \textcolor{gray}{Thoughts: character\_1 wants to get some water, and the possible destination is object\_0-sink. It needs moves area\_1:object\_5-bed $\rightarrow$ area\_2 $\rightarrow$ area\_0:object\_0-sink, according to the scene graph.}

keyframe 2:  
(character\_0, sitting, object\_6-chair, (talk to, character\_1)). \textcolor{gray}{Thoughts: character\_0 goes to the dining area to meet character\_1. It needs moves area\_2:object\_11-sofa $\rightarrow$ area\_2:object\_6\_chair, according to the scene graph.}
(character\_1, sitting, object\_7-chair, (talk to, character\_0)). \textcolor{gray}{Thoughts: character\_1 goes to the dining area to meet character\_0. It needs moves area\_0:object\_0-sink $\rightarrow$ area\_2:object\_7\_chair, according to the scene graph.}
        \end{tcolorbox}

        % \vspace{0.3cm}
        \begin{tcolorbox}[colback=white, colframe=blue, width=\textwidth, boxrule=0.2mm, arc=2mm]
            \textcolor{gray}{Other samples \dots}
        \end{tcolorbox}
    \end{tcolorbox}

    \caption{An example of an MLLM prompting format used in our method. The prompt explicitly guides the model to reason through character activities, spatial transitions, and interactions within the environment.}
    \label{fig:prompt}
\end{figure*}

Building on the foundation established in the previous phase, where MLLMs have gained an understanding of the scene layout and objects inside the scene, we next instruct MLLMs to fill out details of the activities occurring within each keyframe. The objective is to enrich the snapshot $\mathbf{s}_t = \{d\}$ for every keyframe $t$ by generating the characters' actions and interactions within the scene. For all character in a keyframe, the MLLMs are tasked with detailing their descriptions following the defined format. To ensure accurate spatial reasoning, we prompt MLLMs to explicitly describe intermediate reasoning steps when characters move between different areas, referring to the area-based scene graph and following the Layout Chain-of-Thought (L-CoT) prompting approach. An example prompt illustrating this strategy is shown in~\autoref{fig:prompt}.

\subsection{Accommodating Characters in 3D Scenes}
\label{sec:optimization}

Given the generated activity by MLLMs, we move forward to populate the virtual characters into the virtual environments as specified. This step employs an optimization technique akin to that described in ~\cite{li2022interactive,li2023generating}, utilizing simulated annealing~\cite{kirkpatrick1983optimization} with a Metropolis-Hastings state-search step~\cite{metropolis1953equation,hastings1970monte} to optimize the positioning and poses of characters. The exploration of solution space is conducted via Markov chain Monte Carlo (MCMC) sampling, achieving efficiency and near real-time character population. This efficiency is attributed to the predefined positional references and interaction details, which significantly narrows the potential placement range constrained by the generated high-level descriptions.

\begin{figure}[t]
     \centering
     \includegraphics[width=0.48\textwidth]{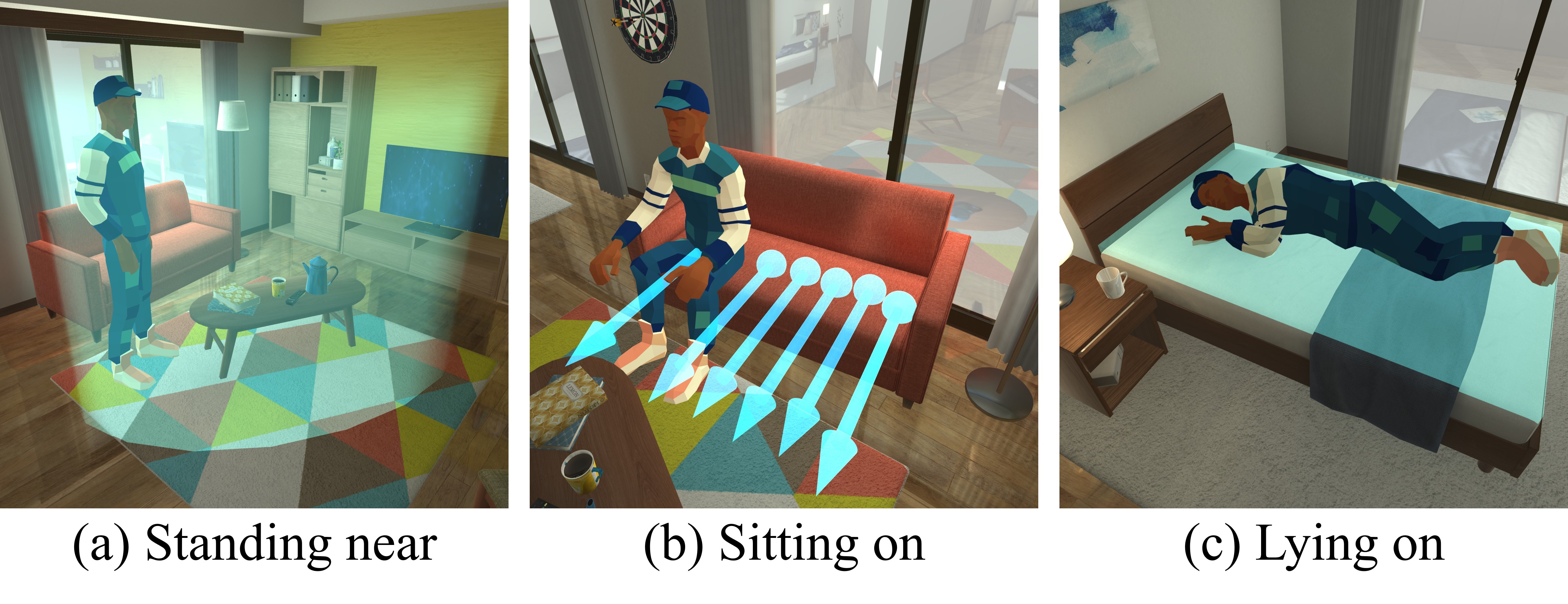}
     \caption{The free spaces considered for different character poses during optimization: (a) \textit{standing near} the furniture, (b) \textit{sitting on} the predicted sitting points, and (c) \textit{lying on} a supporting surface. }
     \label{fig:space}
\end{figure}

A character's pose is parameterized as $(\mathbf{p}, \theta_b, \theta_h)$ conditioned on the fundamental pose, where $\mathbf{p}$ denotes the 3D position, $\theta_b$ the body rotation relative to the positional reference $r$, and $\theta_h$ the head rotation relative to the body, with both simplified rotations occurring around an axis perpendicular to the floor. This distinction between body and head rotations mirrors real-life scenarios, such as a person sitting on a chair who may rotate their head to look at someone or something without significantly rotating their body. The spatial allocation for character placement varies with the pose as illustrated in \autoref{fig:space}: for a \textit{standing} pose, movement is restricted to the proximity of an object; for a \textit{sitting} pose, GoalNet~\cite{hassan2021populating} is used to predict multiple sitting points (i.e., positions and forward directions for the hip joint) and orientations; for a \textit{lying} pose, the character is placed on the largest supporting surface available on appropriate furniture.

The placement process involves specific Markov chain dynamics to propose moves during MCMC sampling, detailed as follows:

\begin{enumerate}
    \item \textbf{Translation}: For a \textit{standing} or \textit{lying} character, the position can be adjusted as $\mathbf{p} \rightarrow \mathbf{p} + \delta \mathbf{p}$, following a bivariate normal distribution. Translation occurs on the xy plane, with height fixed during initialization given the positional reference $r$'s geometry. For the \textit{sitting} pose, a proposed translation moves a character from the occupied sitting point to another nearby.
    \item \textbf{Body Rotation}: A character's body rotation can be adjusted as $\theta_b \rightarrow \theta_b + \delta \theta_b$, following a normal distribution.
    \item \textbf{Head Rotation}: Similarly, the head rotation can be adjusted as $\theta_h \rightarrow \theta_h + \delta \theta_h$ following a normal distribution. For the \textit{lying} pose, the head rotation is fixed as $0$.
\end{enumerate}

\begin{figure*}[t]
     \centering
     \includegraphics[width=\textwidth]{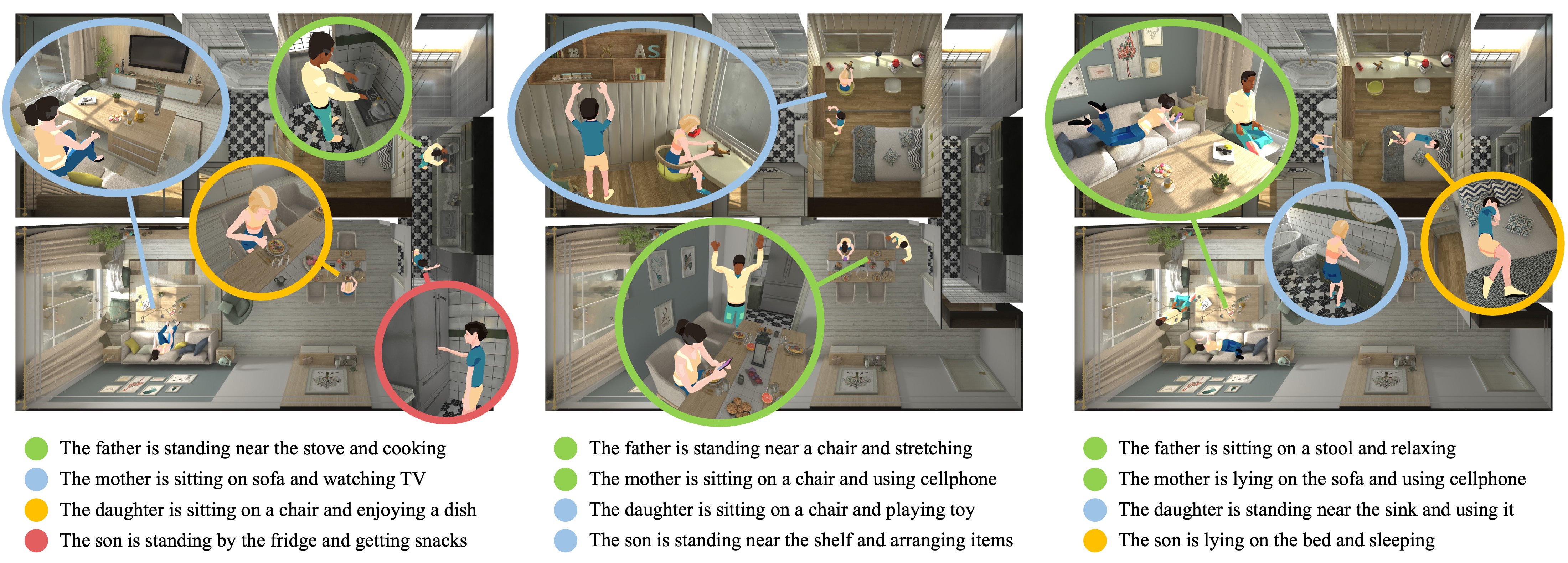}
     \caption{Selected keyframes showcasing an ``apartment'' activities synthesis, arranged sequentially from left to right.}
     \vspace{-2mm}
     \label{fig:apartment}
\end{figure*}
    
Interactions generated by MLLMs are converted into constraints, addressing both positional and rotational requirements for realistic interactions, closely mirroring the method described in ~\cite{li2023generating}. Two constraint templates are defined as:

\begin{enumerate}
    \item \textbf{Positional constraints}, which constrain the distances between a character $a$ and another identity $b$ (a character or and objects). The distance between a character and the positional reference can naturally be satisfied given the definition. However, if the character has physical interactions with certain objects, such distances should be further considered. For example, to use a computer, a desired threshold distance $D$ can be necessary to guarantee the computer is accessible. The template is formulated as:

    \begin{equation}
        C_{\textrm{p}}(a, b) = \text{max}\left(1 - e^{D - \text{distance}(a, b)}, 0\right)
        \label{equ:positional}
    \end{equation}

    \item \textbf{Rotational constraints}, which constrain the direction of a character by measuring the angle between two directions $\vec{x}$ and $\vec{y}$. For example, if a character is looking at another, it can be desired that the facing direction of the character matches the direction towards the other. The template is defined as:

    \begin{equation}
        C_{\textrm{o}}(\vec{x}, \vec{y}) = \frac{1 - \cos(\vec{x}, \vec{y})}{2},
    \end{equation}
\end{enumerate}

We employ an LLM agent to map textual descriptions to constraints. Specifically, the agent translates textual interactions into positional and rotational constraints using the two templates. It also determines appropriate threshold distances $D$ based on its learned common knowledge (e.g. $0.5m$ for interacting with a computer).

Optimization treats these interactions as independent activity groups ${\mathcal{G}}$, where a single character, or a few characters interacting with each other, form an independent group, optimizing their poses jointly. In some cases, characters performing their individual activities can also be grouped if they are assigned to the same positional reference. For example, poses of two characters sitting on the same sofa should be considered together to avoid interpenetration. The optimization within $\mathcal{G}$ uses a Boltzmann-like objective:

\begin{equation}
f(\mathcal{G}) = e^{-\frac{1}{t} C(\mathcal{G})},
\end{equation}

\noindent where $C(\mathcal{G})$ represents the constraints within $\mathcal{G}$ and $t$ is the temperature parameter decreasing over iterations. A proposed move $\mathcal{G}'$ with the Markov chain dynamics is accepted with a probability:

\begin{equation}
\alpha(\mathcal{G}'|\mathcal{G}) = \text{min}\left[\frac{f(\mathcal{G}')}{f(\mathcal{G})}, 1\right] = \text{min}\left[e^{\frac{1}{t}(C(\mathcal{G}) - C(\mathcal{G}'))}, 1\right].
\end{equation}

This pose optimization ensures a realistic portrayal of characters in the virtual environment, aligning with their textual activities and interactions. Animation clips corresponding to the interaction verbs from the generated activity descriptions are applied to the characters. For activities spanning multiple discrete keyframes, if a character moves between different positions, a walking animation is automatically inserted. These movements follow collision-free trajectories computed using the A* algorithm.

\begin{figure*}[htbp]
     \centering
     \includegraphics[width=\textwidth]{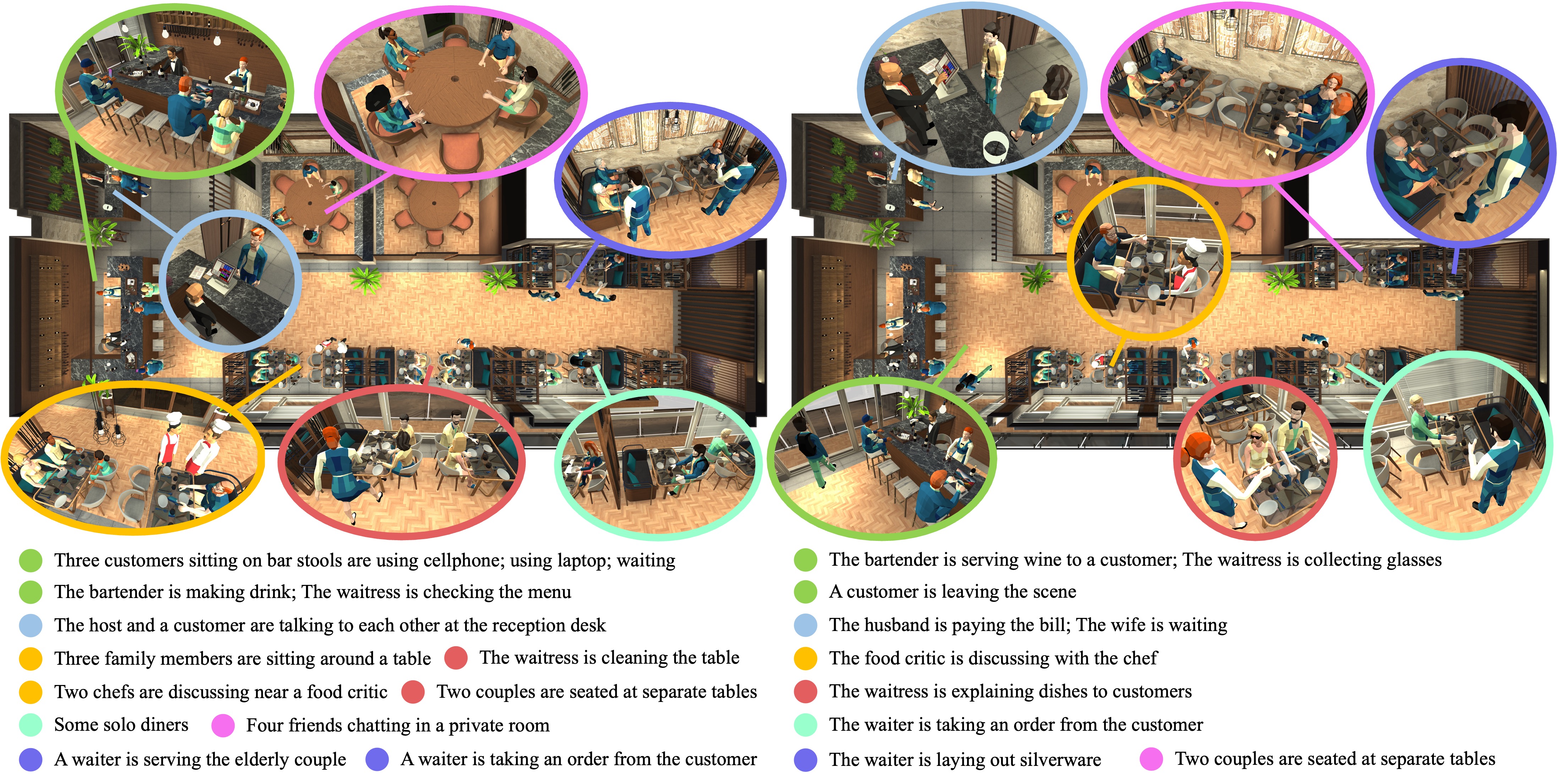}
     \caption{Selected keyframes showcasing an ``restaurant'' activities synthesis, arranged sequentially from left to right.}
     \label{fig:restaurant}
\end{figure*}

\begin{figure*}[htbp]
     \centering
     \includegraphics[width=\textwidth]{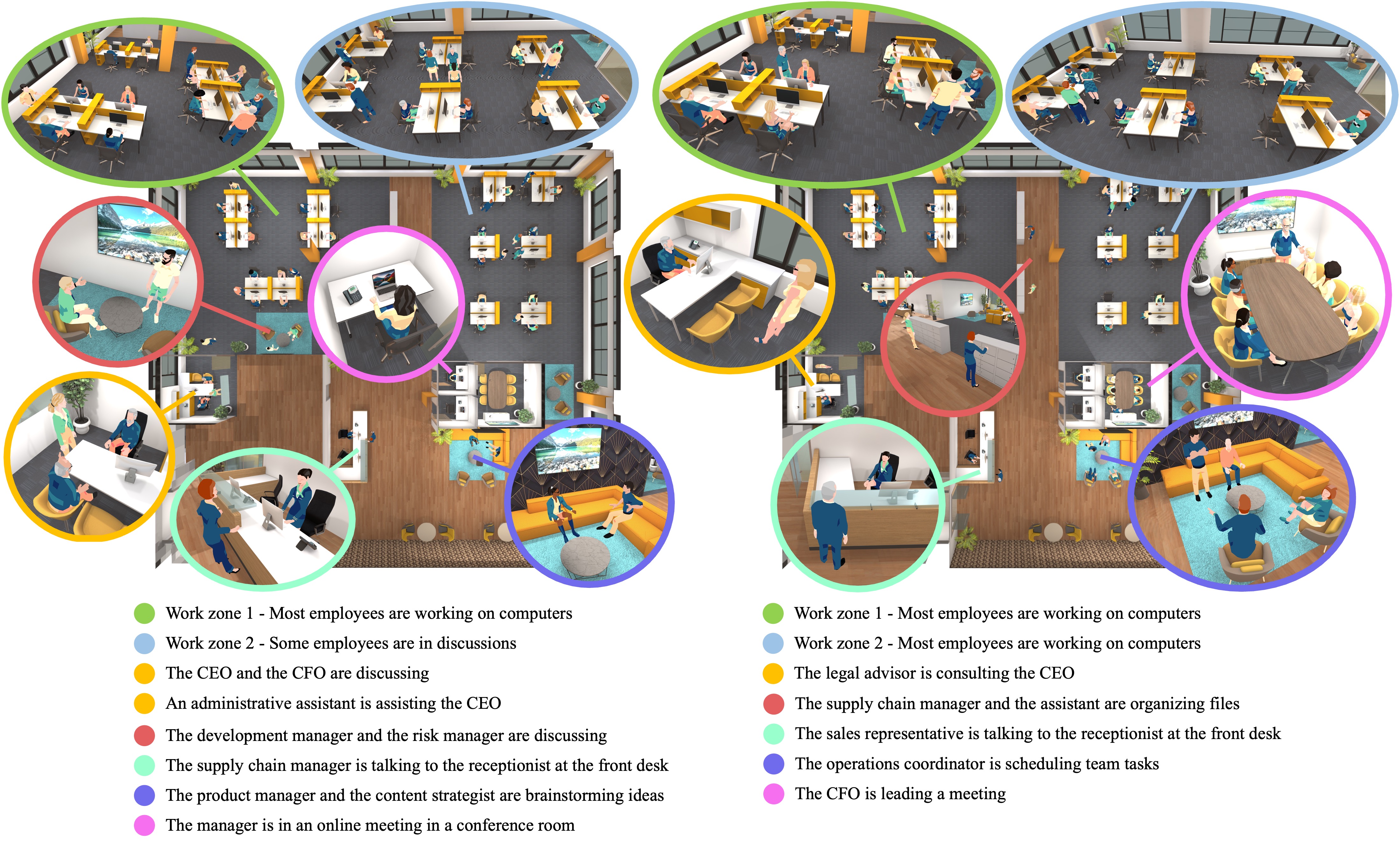}
     \caption{Selected keyframes showcasing an ``office'' activities synthesis, arranged sequentially from left to right.}
     \label{fig:office}
\end{figure*}

\section{Results}

\subsection{Virtual Activity Generation}

We demonstrate the effectiveness of our approach by generating virtual activities across three synthetic scenes of varying scales: a small apartment (4 characters, \autoref{fig:apartment}), a medium restaurant (31 characters, \autoref{fig:restaurant}), and a large office (40 characters, \autoref{fig:office}).

In the apartment scene in~\autoref{fig:apartment}, characters engage in various household activities that reflect typical family life dynamics. A highlighted keyframe exemplifies the details: the father cooking, the mother watching TV, the daughter at the dining table, and the son getting snacks from the refrigerator. The generated poses and actions demonstrate that our method effectively produces contextually coherent and realistic household interactions.

The restaurant scenario in~\autoref{fig:restaurant} showcases our approach’s adaptability by generating a diverse array of character behaviors appropriate for a public dining setting, from dining and interacting with wait staff to engaging in conversations. The accurate depictions of different activities tailored to the restaurant showcase the creativity in handling predefined roles and contextual interactions.

For the large-scale office scenario in~\autoref{fig:office}, our generated results capture a professional environment with characters performing tasks such as working on computers, participating in meetings, or engaging in discussions. The characters are posed in a manner that suggests active engagement with their tasks and surroundings, reflecting the complexity of the generated activities.

\begin{figure}[t]
     \centering
     \includegraphics[width=0.48\textwidth]{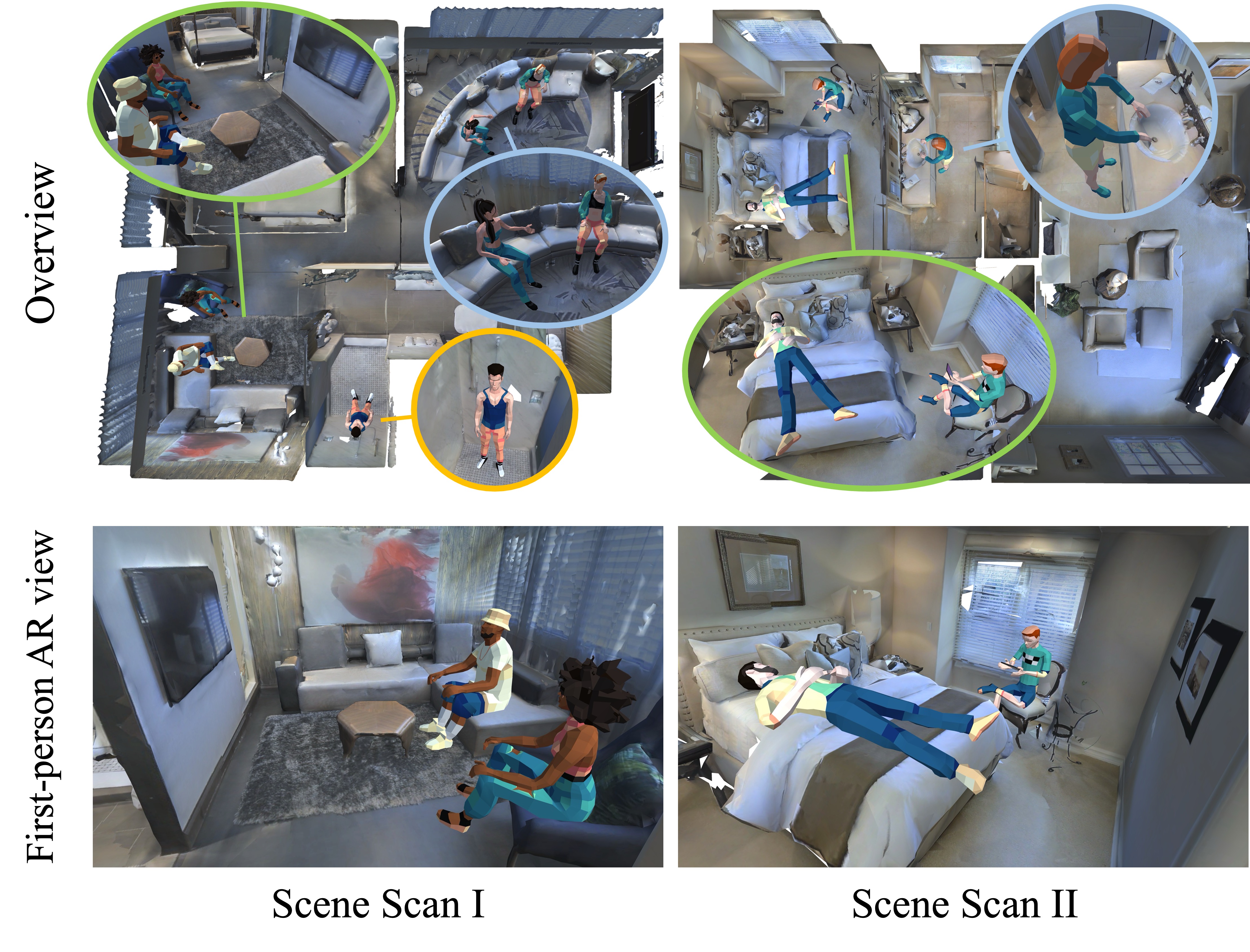}
     \vspace{-8mm}
     \caption{Keyframe snapshots in the 3D scans of two real scenes.}
     \label{fig:scan}
\end{figure}

Additionally, we validate the applicability of our approach to AR scenarios by testing it on real-world 3D scans from Matterport3D dataset~\cite{Matterport3D}. \autoref{fig:scan} shows keyframes from two real-world scenes, highlighting lifelike activities and demonstrating our method’s potential for generating realistic human-scene interactions in practical AR applications, utilizing 3D scans to serve as AR digital twins.

\begin{figure}[t]
     \centering
     \includegraphics[width=0.48\textwidth]{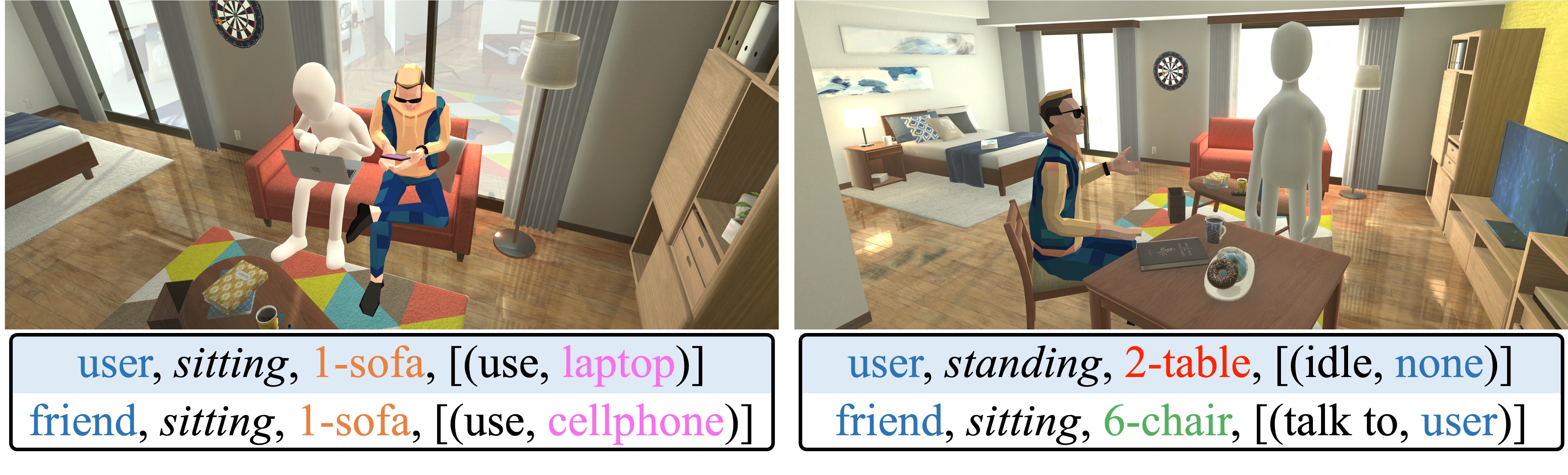}
     \vspace{-6mm}
     \caption{Interactive activity generation in response to user input. When the user avatar (white stickman) takes an action, we fix both the corresponding prompt and 3D pose, then run our method to generate context-aware behaviors for the other characters.}
     \vspace{-3mm}
     \label{fig:dynamic}
\end{figure}

Across all scenes, the attention to spatial dynamics and interaction fidelity is evident. Characters are placed and engaged in a manner that authentically reflects the generated virtual activities, showcasing the adaptability of our method to diverse environments and its effectiveness in creating believable, interactive scenarios. The outcomes validate the generalizability of our approach, demonstrating its scalability and the significant potential of applying MLLMs in enriching virtual experiences through the generation of contextually appropriate and dynamic virtual activities.

Our approach naturally supports dynamic and interactive activity generation. As shown in~\autoref{fig:dynamic}, once a user performs an action, we fix the corresponding prompt to reflect the user’s behavior and run our method to generate responsive behaviors from other characters. The 3D pose optimizer also incorporates the user avatar’s pose as a constraint: For instance, when both the user and a virtual character sit on the same sofa, the character’s placement avoids interpenetration with the user avatar.

The most closely related work to ours is~\cite{li2023generating}, which is, however, not directly comparable. \cite{li2023generating} learns human-scene interactions from video data, making it heavily reliant on the quality of the training data, particularly in terms of the sequence length (number of frames) and the scene scale (number of people and objects). We conducted preliminary tests using it in a restaurant setting, one of the dominant scene types in its training set. In controlled cases with fewer than 8 characters, the generated behaviors were generally reasonable. However, as the scene scale increased, the model began to fail. It exhibited unstable behaviors such as repeatedly adding new characters without meaningful interactions and subsequently removing them in the next keyframe. These issues suggest that~\cite{li2023generating} does not generalize well to larger-scale or more complex scenes without significant additional training. In contrast, our method does not rely on pre-collected interaction sequences or scene-specific training data. Our framework can interpret arbitrary scenes, reason about spatial and social contexts, and generate adaptive multi-character activities at scale. Furthermore, our approach has demonstrated generalizability across diverse scene types without retraining or domain adaptation. We scale up to 40-character scenes with stable behaviors and consistent interaction logic.

\subsection{Ablation Tests}
\label{sec:ablation}

To evaluate the effectiveness of L-CoT and the vision inputs used in our work, we accordingly conducted two ablation tests.

\begin{figure}[t]
     \centering
     \includegraphics[width=0.45\textwidth]{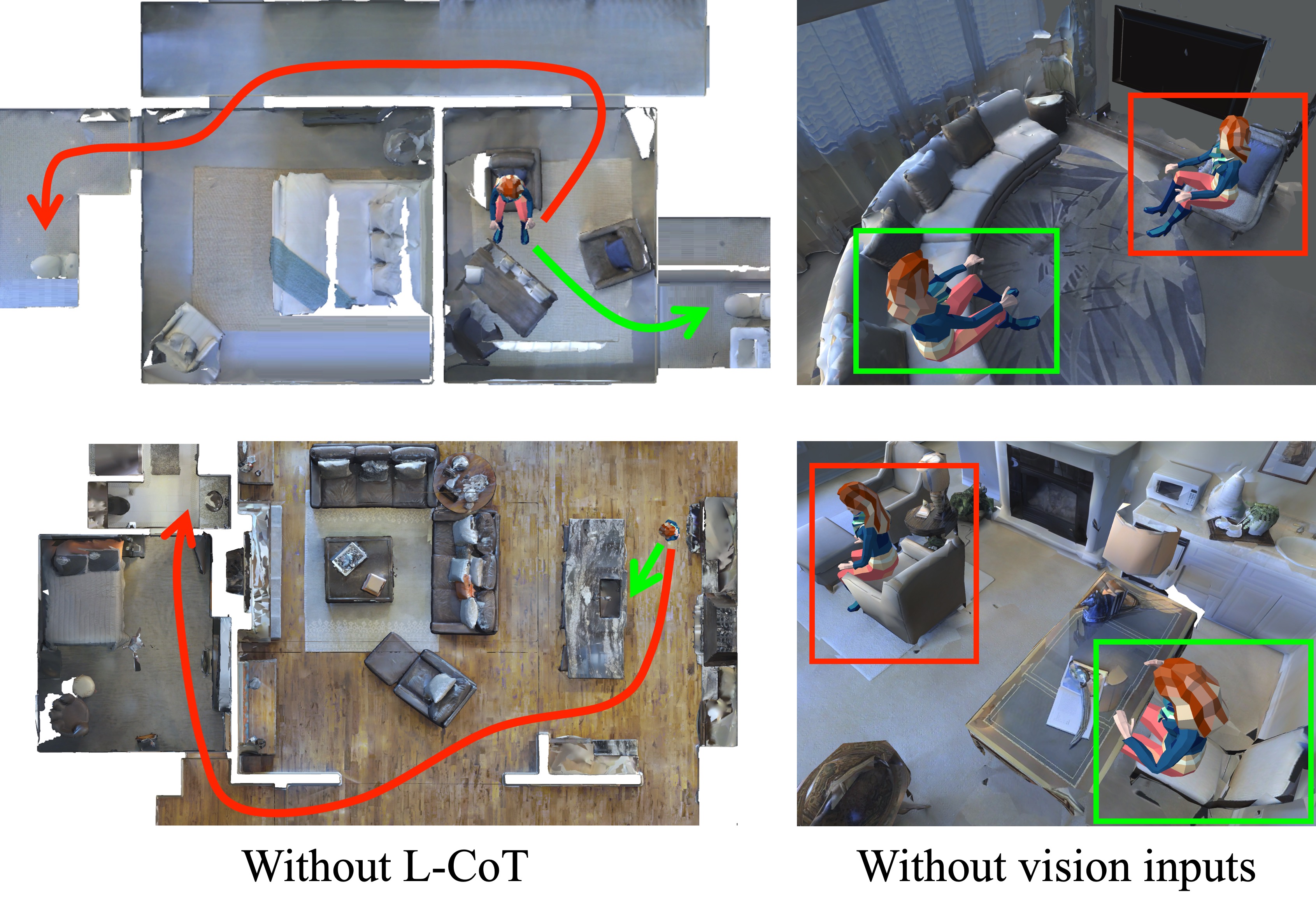}
     \vspace{-4mm}
     \caption{Ablation examples. Left: Without explicit spatial reasoning from L-CoT, the model may select suboptimal, distant destinations (red trajectories) rather than more suitable nearby alternatives (green trajectories).
Right: Without vision inputs, the model may choose inappropriate or incorrectly oriented objects for interactions, resulting in unnatural or inaccessible character poses (red boxes). Incorporating visual guidance ensures contextually appropriate object selection and natural interaction poses (green boxes).}
    \vspace{-3mm}
     \label{fig:ablation}
\end{figure}

We first assessed the activity generation without L-CoT. Given only area descriptions and multi-view scene images, the MLLM can still recognize objects and their visual appearance effectively. However, without guiding it to make explicit spatial reasoning using L-CoT, the MLLM struggled to generate logically consistent activities. As illustrated in~\autoref{fig:ablation} (top-left), when prompted to move the character sitting on a chair to ``use toilet'' next, the model randomly selected one of two bathrooms due to the lack of spatial understanding. Without L-CoT, it occasionally chose a bathroom located inconveniently inside a bedroom (red trajectory). In contrast, the model reliably identified the nearest and most suitable bathroom (green trajectory) with L-CoT. Similarly, in~\autoref{fig:ablation} (bottom-left), without L-CoT, the model inefficiently moved a character, who obtained an apple from the kitchen refrigerator, to ``use sink'' to wash the apple in a distant bathroom (red trajectory), instead of just using the adjacent sink in the kitchen (green trajectory).

We also examined the role of vision inputs in the system. Some recent approaches use scene graph as the only input to LLMs for human activity generation~\cite{gorlo2024long,li2025x}, but rely on carefully designed scene graph representations capturing symbolic concepts such as traversability, objects, and people/agents~\cite{armeni20193d,rosinol2021kimera}. We used scene graphs generated by a standard MST algorithm in this test, providing MLLMs with general spatial layouts but lacked detailed visual information such as object shapes and poses. For instance, as illustrated in~\autoref{fig:ablation} (top-right), when relying solely on the simplified scene graph, the model assigned the character to a chair that resulted in an unnatural and uncomfortable pose (red box) to perform the ``watching tv'' activity. With visual guidance from images, the MLLM avoided inappropriate chairs, resulting in more natural and realistic character poses (green box). Similarly, in~\autoref{fig:ablation} (bottom-right), the character sat on a chair facing away from the desk and positioned farther from it, rendering the desk inaccessible for work (red box). With visual inputs, the model correctly selected a chair closer to the desk and oriented appropriately, enabling natural and functional interactions (green box). While detailed scene graphs designed with comprehensive spatial and object-specific attributes might improve results as well, creating such representations for diverse scene types is challenging. Therefore, directly leveraging visual inputs offers an efficient and effective solution.

\subsection{Perceptual Study}

We conducted a perceptual study, approved by the Institutional Review Boards, to evaluate the quality of virtual activities generated by our method compared to manual professional designs. We invited 5 technical artists, each with over 5 years of professional Unity3D experience, to create virtual activities in 5 different small virtual scenes (scene 1, 2: apartments; scene 3: barbershop; scene 4: bar; scene 5: office). They were informed to include a desired number of virtual characters, and assign roles to all. They were asked to create activities of at least 8 keyframes: in each keyframe, they described all characters' behaviors and states, and adjusted their poses and selected proper animation clips to match the context they wanted. The manual creation mainly included clicking and dragging operations using our Unity3D designer tool.

For each manual design, we summarized all characters' activities using texts, and used these summaries together with characters' roles and the number of keyframes as additional constraints in the prompt, ensuring contextual similarities for fair comparison.

\begin{table}[]
% \vspace{+1mm}
\label{table:perceptual}
\caption{Each cell shows the selection rate ($\text{SR}_{ours}$) of our method over manual designs (percentage ± standard error), along with the $p$-value from a Chi-square test under the null hypothesis of no preference difference. The four evaluation metrics are: reasonableness ($M^r$), naturalness ($M^n$), understandability ($M^u$), and overall ($M^o$). }
\small
\begin{tabular}{ll|l|l|l|l|l}
\hline
\multicolumn{2}{l|}{Metrics}                         & Scene 1       & Scene 2       & Scene 3       & Scene 4       & Scene 5       \\ \hline
\multicolumn{1}{l|}{\multirow{2}{*}{$M^r$}} & $\text{SR}_{ours}$ & 51±5.00 & 48±5.00 & 49±5.00 & 43±4.95 & 54±4.98 \\
\multicolumn{1}{l|}{}                      & \textit{p}-value       & 0.84    & 0.69    & 0.84    & 0.16    & 0.42    \\ \hline
\multicolumn{1}{l|}{\multirow{2}{*}{$M^n$}} & $\text{SR}_{ours}$ & 53±4.99 & 55±4.97 & 56±4.96 & 48±5.00 & 44±4.96 \\
\multicolumn{1}{l|}{}                      & \textit{p}-value      & 0.55    & 0.32    & 0.23    & 0.69    & 0.23    \\ \hline
\multicolumn{1}{l|}{\multirow{2}{*}{$M^u$}} & $\text{SR}_{ours}$ & 47±4.99 & 49±5.00 & 53±4.99 & 45±4.97 & 51±5.00 \\
\multicolumn{1}{l|}{}                      & \textit{p}-value      & 0.55    & 0.84    & 0.55    & 0.32    & 0.84    \\ \hline
\multicolumn{1}{l|}{\multirow{2}{*}{$M^o$}} & $\text{SR}_{ours}$ & 45±4.97 & 54±4.98 & 59±4.92 & 52±5.00 & 47±4.99 \\
\multicolumn{1}{l|}{}                      & \textit{p}-value      & 0.32    & 0.42    & 0.07    & 0.69    & 0.55    \\ \hline
\end{tabular}
% \vspace{-4mm}
\end{table}

We recruited 100 participants in Amazon
Mechanical Turk, 77 males and 23 females with ages ranging from 25 to 52 years, for this study. We animated both manually created and automatically generated activities, and rendered video recordings for all. For each participant, we showed them all 5 scenes, each consisting of a result under \textit{designer} condition and another under \textit{ours} condition. The study followed a two-alternative forced choice (2AFC) format: for each paired video recordings (randomly shuffled), they watched and chose a better one regarding each of the four evaluation metrics:

\begin{itemize}
    \item Reasonableness ($M^r$): How reasonable and contextually appropriate are the characters’ actions and poses, given the scene and the described activity.
    \item Naturalness ($M^n$): How natural, and lifelike are the characters’ movements and body poses.
    \item Understandability ($M^u$): How easy is it to understand what the characters are doing by watching the video.
    \item Overall ($M^o$): General preference considering all aspects.
\end{itemize}

The results in ~\autoref{table:perceptual} shows the numerical results. The selection rates for \textit{ours} ($\text{SR}_{ours}$) were generally close to $50\%$, indicating similar preferences to \textit{designer} results.  A Chi-square test confirmed that the differences were not statistically significant ($p > 0.05$), supporting the conclusion that our method produces results of comparable perceptual quality to human-designed content. 

For efficiency, professional designers spent 13–32 minutes per scene. Our method completed all scenes in 3–5 minutes, including image processing, MLLM generation, and 3D pose optimization.

\section{Limitations And Future Work}

While our method demonstrates promising results, several limitations remain. MLLMs could still struggle when processing scenes or activities with extended contexts, such as large environments or activities of long keyframe sequences. For instance, in our office scene, although the generated activities mostly matched the environment, certain social conventions, such as employees having assigned desks, were overlooked. We found that generating fewer keyframes per query, rather than requesting multiple keyframes at once, improves context consistency and robustness. It is worth further exploring hierarchical reasoning methods, and incorporating explicit knowledge about social or professional norms, which might allow MLLMs to maintain context coherence over longer activity sequences and across larger spaces.

Additionally, our current method relies on predefined animation clips tied directly to interaction verbs described by the MLLMs, which may lead to repetitive animations across different characters performing similar actions. Although it can be easily extended by adding more animations and interaction verbs to support richer, more diverse behaviors, another promising improvement would be integrating generative motions~\cite{min2009interactive,holden2016deep,qin2022motion,huang2023real} to allow smoother, more diverse, and adaptive animations that reflect detailed variations between characters and interactions.

Our method separates the process into two stages: activity reasoning with an MLLM and 3D pose optimization. This effective modular design opens up an interesting direction for future research—enabling MLLMs to directly generate human poses. Achieving this goal would require aligning the language modality with parametric human models such as SMPL~\cite{loper2015smpl}. Recent work has begun exploring the use of MLLMs to generate SMPL parameters directly~\cite{feng2024chatpose,li2024unipose,lin2024chathuman,delmas2024poseembroider}, reflecting growing interests in this area. However, current approaches remain limited in their ability to model complex multi-character interactions and scene-aware behaviors. Extending MLLMs to reason about and generate physically and socially plausible multi-character poses remains a challenging yet promising direction for future work.

\section{Conclusion}

In this work, we propose a novel framework that enriches virtual environments with realistically virtual activities, leveraging the advanced capabilities of multimodal large language models (MLLMs). By integrating vision-language modalities with Layout Chain-of-Though (L-CoT) prompting, our approach enables MLLMs to interpret spatial and semantic cues from virtual scenes and generate context-aware, detailed descriptions of adaptive activities. These descriptions, rich in contextual understanding, lay the groundwork for the precise characters' placements within the virtual space. Our approach ensures that characters are not just accurately placed but are also seemingly engaged in believable interactions, thanks to a robust optimization framework that aligns with the structured format of activity descriptions.

The adaptation of MLLMs for virtual activity generation underscores their potential in crafting complex, dynamic, and immersive scenarios. The implications of our findings suggest promising avenues for enhancing user experiences for applications in VR/AR, gaming and AI-driven storytelling. The adaptability and scalability of our method suggest its applicability across a broad spectrum of virtual environment creations such as future metaverse experiences.

\acknowledgments{
This project is supported by National Science Foundation grants with award numbers 2430673, 2418236, and 1942531.}

\bibliographystyle{abbrv-doi}

\bibliography{bib}
\end{document}